\documentclass[aps,prl,reprint,superscriptaddress,showpacs,nofootinbib]{revtex4-1}

\usepackage{graphicx}
\usepackage{color}
\usepackage{amsfonts}

\usepackage{hyperref}

\usepackage{ulem}

\begin{document}

\title{Ultra-relativistic electron beams deflection by quasi-mosaic crystals}

\author{Gennady B. Sushko}
\email[]{sushko@mbnexplorer.com}
\affiliation{MBN Research Center, Altenh\"{o}ferallee 3, 60438 Frankfurt am Main, Germany}

\author{Andrei V. Korol}
\email[]{korol@mbnexplorer.com}
\altaffiliation{On leave from: St. Petersburg State Marine Technical University, Leninsky ave. 101,
198262 St. Petersburg, Russia}
\affiliation{MBN Research Center, Altenh\"{o}ferallee 3, 60438 Frankfurt am Main, Germany}

\author{Andrey V. Solov'yov}
\email[]{solovyov@mbnresearch.com}
\altaffiliation{On leave from:
Ioffe Physical-Technical Institute, Politekhnicheskaya 26, 194021 St. Petersburg, Russia}
\affiliation{MBN Research Center, Altenh\"{o}ferallee 3, 60438 Frankfurt am Main, Germany}

\begin{abstract}
This paper  provides an explanation of the key effects behind the deflection of ultra-relativistic electron beams
by means of oriented ‘quasi-mosaic’ Bent Crystals (qmBC). 
It is demonstrated that accounting for specific geometry of the qmBC and its orientation
with respect to a collimated electron beam, its size and emittance is essential for an accurate
quantitative description of experimental results on the beam deflection by such crystals. 
In an exemplary case study a detailed analysis of the recent experiment at the SLAC facility is presented.
The methodology developed has enabled to understand the peculiarities in the measured distributions of
the deflected electrons.
Also, this achievement constitutes an important progress in the efforts towards the practical 
realization of novel gamma-ray 
crystal-based light sources and puts new challenges for the theory and experiment in this research area.
\end{abstract}


\maketitle

In recent years significant efforts of 
the research and technological communities
have been devoted to design and practical realization of novel gamma-ray Crystal-based Light Sources
(CLS) that can be set up by exposing oriented linear, bent or periodically bent crystals  to
beams of ultrarelativistic positrons or electrons \cite{KorolSolovyovCLS2020,KorolSolovyovColloquium2021}. 
Brilliance of radiation emitted in a crystalline undulator LS by available beams in the
photon energy range  $10^0$-$10^1$  MeV, being inaccessible to conventional synchrotrons, undulators and
XFELs, greatly exceeds that of laser-Compton scattering LSs and is higher than predicted in the
Gamma Factory proposal to CERN \cite{Krasny2018}. 
Manufacturing of CLSs will have significant impact on many research areas in physics, chemistry, 
biology, material science, technology and medicine, being a subject
of current European projects 'N-LIGHT' \cite{NLight} 
and TECHNO-CLS \cite{TECHNOCLS}.

So far oriented crystals exposed to beams of charged particles have been already  
utilised in a number of applications for beams manipulation, such as steering, bending, extraction and focusing, 
see \cite{KorolSolovyovColloquium2021,ChannelingBook2014} 
and references therein.
These and other newly emerging applications in this research area require high quality crystals
(bent or periodically bent) and
collimated beams of charged ultrarelativistic particles of different energies.

Construction of novel CLSs is a challenging task involving a broad range of correlated research and technological 
activities \cite{KorolSolovyovCLS2020,KorolSolovyovColloquium2021}.
During the last decade a number of papers published in  high-impact journals
\cite{MazzolariPRL2014,Wienands2015,BandieraPRL2015,BandieraPRL2013,MazzolariEPJC2018,SytovEPJC2016,WistisenPRAB2016}
on channeling and channeling radiation experiments with bent crystals at different facilities (SLAC, CERN, MAMI).
This paper reports on the important progress in this field providing an explanation
of the key effects arising by  deflection of  ultrarelativistic electron and positron  beams
by means of oriented ‘quasi-mosaic’ Bent Crystals (qmBC). 
It is demonstrated that account for specific geometry of qmBC and its orientation
with respect to a collimated beam of projectile particles, the beam size and emittance 
is essential for the quantitative description of the experimental results 
on the beam deflection by such crystals. 

Manufacturing of crystals of different desired geometry is an important
technological task in the context of their applications in the gamma-ray CLSs and
the aforementioned experiments.
The systematic review of different technologies
exploited for manufacturing of crystals of different type, geometry, size, quality, etc
is given in \cite{ChannelingBook2014,KorolSolovyovCLS2020,KorolSolovyovColloquium2021}. 
A short summary of several relevant approaches that have been utilized to produce bent crystals is provided
in Supplemental Material (SM).

The high-quality qmBCs  structures with desirable and fully controllable parameters have been
manufactured  for the aforementioned channeling experiments by the following means 
\cite{GuidiJPD2009,CamattariEtAl2015,MazzolariEtAlEPJCv882018}.
When a moment of force is applied to a crystalline material, some
secondary curvatures may arise within the solid \cite{Lekhnitskii1981}.
A well known secondary deformation is the anticlastic curvature 
with radius $R_{\rm a}$
that occurs in a medium subjected to two moments.
In particular, it occurs in the perpendicular direction with respect to the primary curvature.
When the two curvatures are combined, the deformed crystal acquires the shape of a saddle.
In contrast to an amorphous medium physical properties of crystals may be strongly anisotropic.
Another type of the deformation caused by anisotropic effects is the ‘quasi-mosaic’ (QM) curvature 
\cite{SumbaevJETP1957,IvanovJETPLett2005}.
QM bent crystals belong to a class of bent crystals featuring two curvatures of
two orthogonal crystallographic planes.

\begin{figure}[h]
\includegraphics[scale=0.14,clip]{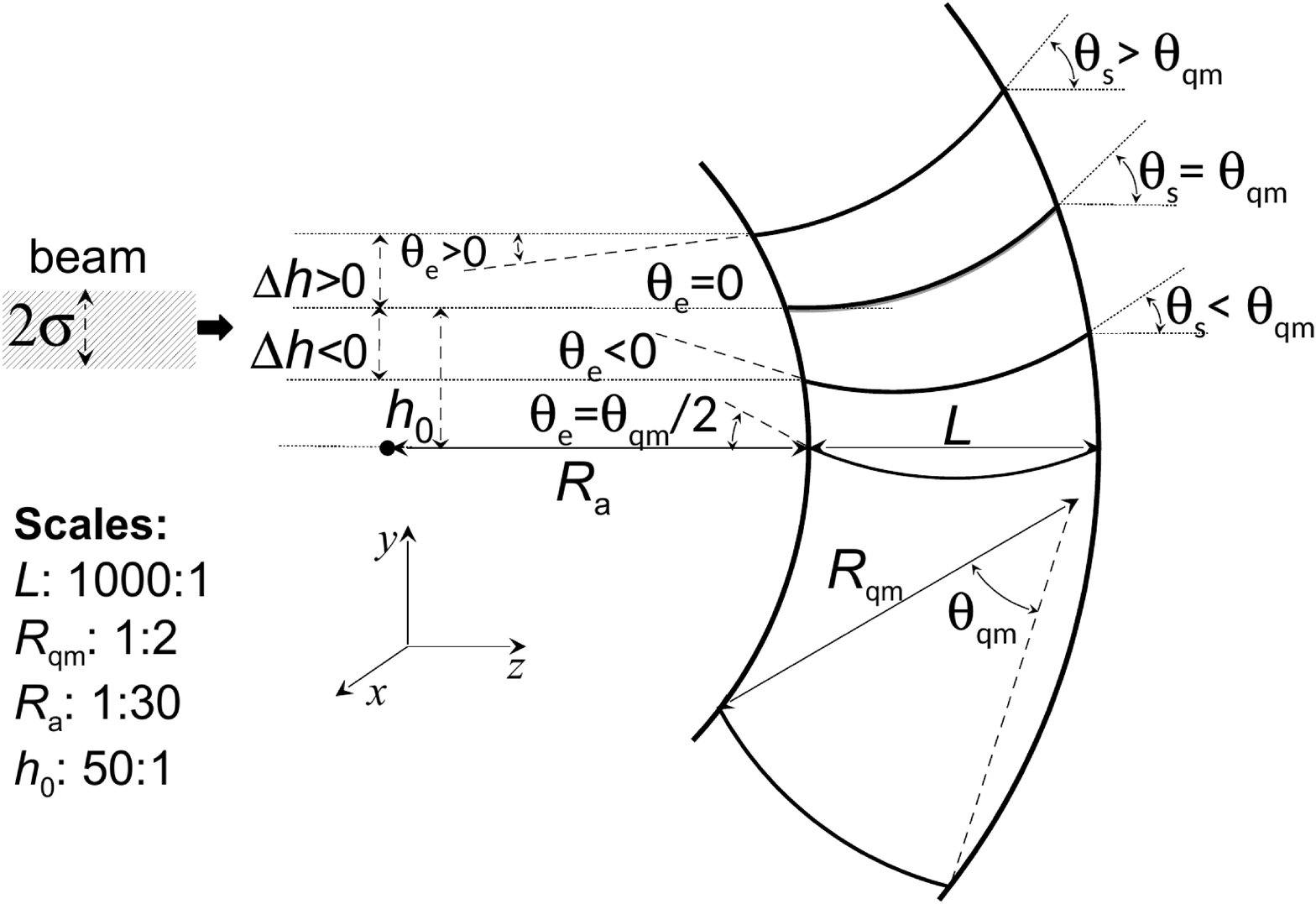}
\caption{
Geometry of the anticlastic and QM bending of a crystal plate of thickness $L$ 
and its orientation with respect to an incident beam  
(shaded rectangle).
The crystal thickness, the anticlastic $R_{\rm a}$ and QM $R_{\rm qm}$ radii
shown in the picture are scaled to meet the values 
indicated in Refs. \cite{GuidiJPD2009,Wienands2015}.
In the experiment \cite{Wienands2015} the $y$ direction was chosen along the 
$\langle 111 \rangle$ axis.
Further explanations are given in the text.
}
\label{Figure01.fig}
\end{figure}

In order to understand the effects arising during channeling of charged particles through qmBC
one should consider the geometry of such a crystal and its orientation with respect 
to an incident beam.
This geometry is shown in Fig. \ref{Figure01.fig}.
For the sake of clarity the case of planar channeling is addressed below. 

Consider a crystal whose planes, which are parallel to the $(xy)$ plane, 
experience anticlastic bending with the curvature radius $R_{\rm a}$.
The center $O$ of the curvature lies on the $z$ axis, which runs through the crystal center.
The QM bending deforms the crystal planes parallel to the $(xz)$ plane.
In what follows it is assumed that $R_{\rm a}$ and the QM bending radius $R_{\rm qm}$ greatly 
exceed the crystal thickness $L$.
These conditions were met for the qmBC samples used in the experiments 
\cite{MazzolariPRL2014,Wienands2015,BandieraPRL2015,BandieraPRL2013,MazzolariEPJC2018,SytovEPJC2016, %
WistisenPRAB2016}.
The QM bending angle is defined as follows
\begin{eqnarray}
\theta_{\rm qm} = {L / R_{\rm qm}} \ll 1\,.
\label{Geometry:eq.06a}
\end{eqnarray}

To start with, let us assume an ideally collimated narrow beam (i.e. that of zero divergence and 
zero beam size in the $y$ direction, $\sigma_{\phi}, \sigma \to 0$) incident on the crystal 
along the $z$ direction.
For a planar channeling 
the beam size and divergence in the $x$ direction do not play important role and thus are 
not considered below.

At the crystal entrance, the angle $\theta_{\rm e}$ between the beam direction and a tangent line to the
QM bent plane depends on the beam displacement $h$ along the $y$-axis:
\begin{eqnarray}
\theta_{\rm e}(h) = {h / R_{\rm a}} - {\theta_{\rm qm} / 2} = {\Delta h / R_{\rm a}}
\label{Geometry:eq.06e}
\end{eqnarray}
where $\Delta h = h - h_0$ with
\begin{eqnarray}
h_0 = \theta_{\rm qm} R_{\rm a} / 2
\label{Geometry:eq.07}
\end{eqnarray}
being the displacement for which the entrance angle $\theta_{\rm e}=0$, i.e.
the tangent line is parallel to the $z$ axis.

A probability of a particle to be accepted into the channeling mode becomes significant if 
$\theta_{\rm e}$ does not exceed Lindhard's critical angle $\theta_{\rm L}$.
Then, using (\ref{Geometry:eq.06e}) one finds the maximum value of $\Delta h$
\begin{eqnarray}
\Delta h_{\max} = \theta_{\rm L} R_{\rm a}\, ,
\label{Distribution:eq.01}
 \end{eqnarray}
so that the channeling condition is met for the particles with
$h$ within the interval $h_0 \pm \Delta h_{\max}$.

At the crystal exit, the angle $\theta_{\rm s}$ between the tangent line and
the beam direction is related to $h$ via
\begin{eqnarray}
\theta_{\rm s}(h) = \theta_{\rm e}(h) + \theta_{\rm qm}\,.
\label{Distribution:eq.05}  
\end{eqnarray}
Hence, the projectiles that are accepted at $y=h$ and channel
through the whole crystal are deflected by the angle lying within the interval
$\theta_{\rm s}(h) \pm \theta_{\rm L}$.

The particles that enter having $\Delta h < 0$ can experience either volume capture 
 or volume reflection \cite{TaratinVorobievPLA1986,TaratinVorobievPLA1987}
in the crystal.  
The geometry analysis for these regimes is given in SM.
The particles that enter with $\Delta h > \Delta h_{\max}$ are neither 
accepted nor experience the volume reflection
but experience multiple scattering which becomes closer to the scattering 
in the amorphous medium as $\Delta h$ increases.

Consider now a Gaussian beam, with width $\sigma>0$ and divergence $\sigma_{\phi}>0$, 
that is incident on the crystal being centered at $y=h$. 
For a beam centered at $h$ most of its particles enter the crystal having the 
transverse coordinates lying within the interval from $h-\sigma$ to $h+\sigma$ and the corresponding 
incident angles $\theta_e$. 
Therefore, the distribution of deflected particles becomes a superposition of different 
propagation scenarios discussed above.

Below in the paper we demonstrate that it is important to know the values of $\sigma$ and $\sigma_{\phi}$ 
as well as of $R _{\rm a}$ quite accurately  
to be able to interprete results of the experiments on beam propagation through oriented qmBC crystals.

In what follows we focus on the analysis of the experiment at SLAC \cite{Wienands2015},
although the physics discussed and the conclusions drawn are applicable to other aforementioned experiments 
with oriented qmBC.
In the experiment, a 60 $\mu$m thick Si(111) qmBC was exposed to a 6.3 GeV electron beam. 
To deduce the values of $\sigma$ and $\sigma_{\phi}$ one can rely on the following 
description provided in the cited paper:
(i) "\dots a beam width of $< 150$ $\mu$m (1$\sigma$) in the vertical and horizontal plane",
and 
(ii) "The beam divergence was inferred \dots to be less than 10 $\mu$rad". 
The QM bending radius of the (111) planes was quoted as $R_{\rm qm} = 15$ cm.
It was mentioned that some measures had been taken "to reduce the anticlastic deformation" 
although the explicit value of $R_{\rm a}$ was not indicated.
Indirectly, one can estimate $R_{\rm a}$ basing on the data presented in \cite{GuidiJPD2009}.
This paper, cited in Ref. \cite{Wienands2015},
discusses the QM bending of Si(211), i.e.
it refers to a  different geometry in which the (111) planes experience the anticlastic bending
rather than the QM one.
For this geometry the value $R_{\rm a}=366$ cm on the centre of the sample was measured.
In our simulations we considered $R_{\rm a}$ 
as a parameter varied within the interval $100-300$ cm. 
Using the aforementioned value of $R_{\rm qm}$ in (\ref{Geometry:eq.06a}) 
one finds $\theta_{\rm qm} = 400$ $\mu$rad. 
Fixing $R_{\rm a}$ and taking into account that 
for a $6.3$ GeV electron Lindhard's critical angle is $80$ 
$\mu$rad \cite{Wienands2015}
one calculates $h_0$ and the maximum displacement $\Delta h_{\max}$.

Numerical modeling of the channeling and related phenomena beyond the continuous potential framework
can be carried out by means of the multi-purpose software package \textsc{MBN Explorer} 
\cite{MBNExplorer2012,MBNChannelingPaper2013,MBNExplorerBook} and  a supplementary special multitask 
software toolkit \textsc{MBN Studio} \cite{MBNStudio2019}.
The  \textsc{MBN Explorer}  was originally developed as a universal computer program to allow
multiscale simulations of structure and dynamics of molecular systems.

\textsc{MBN Explorer} simulates  the motion of relativistic projectiles along with
dynamical simulations of the crystalline environment \cite{MBNChannelingPaper2013}.
The computation accounts for the interaction of projectiles with separate atoms of the
environment, whereas a variety of interatomic potentials implemented supports
rigorous simulations of various media.
Overview of the results on channeling and radiation of charged particles in inear, bent and 
periodically bent crystals
simulated by means of  \textsc{MBN Explorer} can be found in 
\cite{KorolSolovyovCLS2020,KorolSolovyovColloquium2021,ChannelingBook2014,MBNExplorerBook}.

To model propagation of particles through qmBCs 
further development of the algorithm for the atomistic simulations of the crystalline media has 
been performed in this work.
The implemented algorithm enabled simulations of a qmBC defined through a transformation of the 
unperturbed crystalline medium by three curvatures (primary, anticlastic and QM), 
positioning of the qmBC 
with respect to the beam direction and the relativistic molecular dynamics
in such environment. 
The results reported below have been obtained by means of this newly implemented algorithm.

\begin{figure}[h]
\includegraphics[clip,scale=0.32]{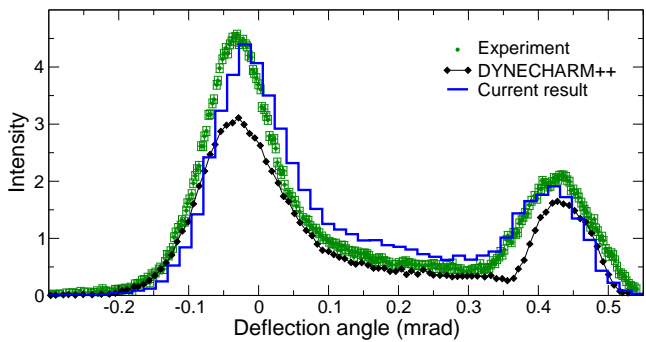}
\caption{
Simulated distribution (histogram)
for the electron beam with size $\sigma=75$ $\mu$m and divergence $\sigma_{\phi}=10$ $\mu$rad
centered at $h=675$ $\mu$m at the entrance of the qmBC with 
$R_{\rm a}=300$ cm.
Open circles with errorbars stand for the experimental data \cite{Wienands2015}.
Both dependences are normalized to the unit area.
Diamonds represent the DYNECHARM++ \cite{BagliGuidiNIMBv3092013}
simulations as they are shown in figure 3 
in \cite{Wienands2015}.
}
\label{Figure02.fig} 
\end{figure}

The main outcome of numerical analysis carried out in this Letter 
in connection with the SLAC experiment is shown in Figure \ref{Figure02.fig}, 
which compares the current simulations with the experimentally measured 
intensity of the deflected electron beam as well as with the 
result of the DYNECHARM++ simulations.
The latter intensities were obtained by digitalizing the data, 
which are presented in arbitrary units in Fig. 3  in \cite{Wienands2015}, 
followed by the background (ca 1.4 a.u.) subtraction.
The resulting experimental values were rescaled to provide the unit area within the interval 
$-0.3\dots0.55$ mrad of the deflection angle.
The ratio experiment-to-DYNECHARM++ was kept as in the original figure.

The simulated and measured angular distributions have the characteristic pattern of the two 
well pronounced peaks interlinked by an intermediate region.
The left peak in the vicinity of $\theta_{\rm s}=0$  describes a fraction of particles
propagating though the qmBC in the forward direction. 
These particles experience multiple scattering resulting in broadening of the
initial distribution of the beam particles. 
Small shift of the peak towards negative angles is due to the volume reflection of 
the particles from the bent planes.
As discussed in SM this effect becomes more pronounced at the entrance points within the region
$-h_0 < h < h_0$.  
The right peak is formed by the particles accepted to the channeling 
regime at the entrance and deflected to the angle $\theta_{\rm s}(h)$ according to
Eq. (\ref{Distribution:eq.05}). 
Our simulations have shown that the position of the channeling peak is determined by the value 
$h$ corresponding to the beam center at the 
entrance point and the width of the peak is determined by the distribution of 
$\theta_{\rm e}(h)$ for the particles of the beam and by Lindhard's angle. 
The peak is also influenced by the  dechanneling process that
is responsible for the formation of the distribution of the deflected particles 
in the region  between the two peaks. 

As mentioned, the angular distribution is very sensitive to the choice of the beam size 
$\sigma$, bending radius $R_{\rm a}$ and the entrance coordinate $h$.
The current simulations
presented in Fig. \ref{Figure02.fig} correspond to a particular set of
these parameters: $\sigma=75$ $\mu$m, $R_{\rm a}=300$ cm and $h=675$ $\mu$m.
It has been established that these values provide close agreement with the 
experimentally measured distribution.
We noted that in Ref. \cite{Wienands2015} the exact value of $\sigma$  
has been specified whereas the values of $R_{\rm a}$ and $h$ as well as their
impact on the profile of the distribution have not been mentioned at all.
Same refers to the results of the DYNECHARM++ simulations.

\begin{figure*}[ht]
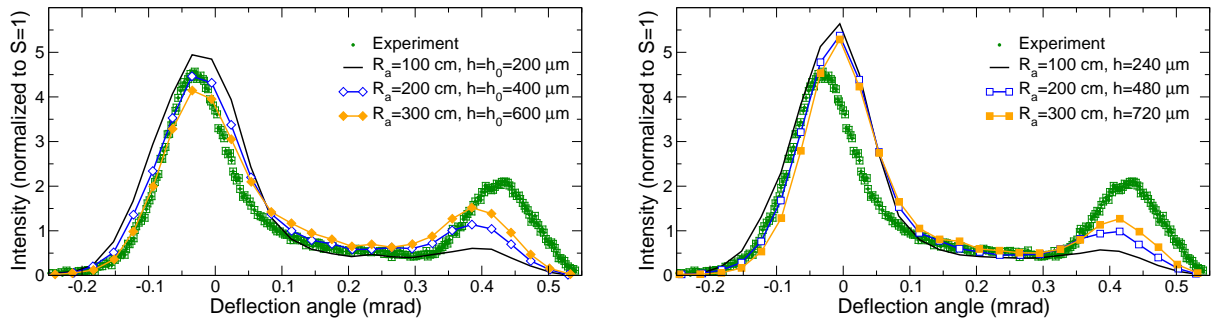

\includegraphics[clip,scale=0.3]{Figure03b.eps}
\hspace*{0.5cm}
\includegraphics[clip,scale=0.3]{Figure03a.eps}
\caption{
Simulated distributions obtained for the beam size $\sigma=150$ $\mu$m  and 
divergence $\sigma_{\phi}=10$ $\mu$rad
but different values of the displacement $h$ and anticlastic radius $R_{\rm a}$.
\textit{Left} panel refers to $h=h_0$.
The $h$ values indicated in the \textit{right} panel   
correspond to $\theta_{\rm s}(h) = 0.44$ mrad. 
All dependences are normalized to unit area.
See also explanation in the text and Fig. S2 in SM.
}
\label{Figure03.fig}
\end{figure*}

\begin{figure*}[h]
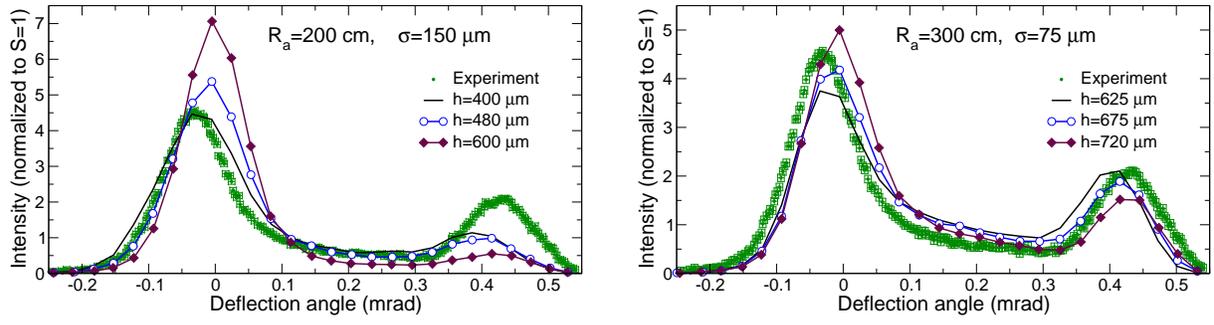

\includegraphics[clip,scale=0.3]{Figure04a.eps}
\hspace*{0.5cm}
\includegraphics[clip,scale=0.3]{Figure04b.eps}
\caption{
Simulated distributions (solid lines with and without symbols) of the
deflected beam obtained for different values of the displacement $h$.
\textit{Left} panel corresponds to the anticlastic radius 
$R_{\rm a}=200$ cm and beam size $\sigma=150$ $\mu$m;
\textit{Right} panel - to $R_{\rm a}=300$ cm and  $\sigma=75$ $\mu$m.
Symbols with error bars show the experimental
distribution \cite{Wienands2015}.
}
\label{Figure04.fig}
\end{figure*}

Figures \ref{Figure03.fig} and \ref{Figure04.fig} illustrate the impact of 
variation of $\sigma$, $R_{\rm a}$ and $h$ on the 
the angular distribution.
The symbols with error bars stand for the experimental 
data obtained as described above.

Figure \ref{Figure03.fig} shows the distribution for a beam with $\sigma=150$ $\mu$m  
incident on the crystal bent with different anticlastic radius as indicated.
In the left panel, each simulation refers to the beam centered at
$h=h_0$ and thus most of the accepted particles are deflected by the angle $\theta_{\rm qm}$
resulting in the channeling peak centered at about 0.40 mrad, which is less than in 
the experiment (ca 0.44 mrad).
The peak intensity increases with $R_{\rm a}$ in accordance with the geometrical arguments 
discussed above. 
Indeed, for $R_{\rm a}=100$ cm the maximum displacement $\Delta h_{\max}=80$ $\mu$m is nearly 
two times less than $\sigma$ resulting in a small fraction of the accepted particles.
Since $\Delta h_{\max} \propto R_{\rm a}$ (see Eq. (\ref{Distribution:eq.01}) and Fig. S2 in SM) 
then for $R_{\rm a}=300$ cm the value of $\Delta h_{\max}$ exceeds $\sigma$ leading to 
the higher intensity. 
The qmBC geometry provides also a qualitative explanation of the changes occurring 
to the left peak. 
For the smallest radius, the inequality $\Delta h_{\max}<\sigma$ suggests that 
large number of particles enters the crystal having the 
transverse coordinate (i) larger than $h_0+\Delta h_{\max}$, and (ii) lower than $h_0-\Delta h_{\max}$.
The former particles contribute mainly to the amorphous-like distribution 
whereas the latter ones can undergo the volume reflection giving rise to the intensity at 
$\theta_{\rm s}<0$.
As $R_{\rm a}$ increases the numbers particles of both types decreases making the peak
narrower and less intensive.

Aiming at bringing the channeling peak position closer to the measured one
another run of simulations has been performed with the same values of $\sigma$ and $R_{\rm a}$  
but different set of initial coordinates of the beam center.
The distributions shown in Fig. \ref{Figure03.fig} \textit{right} refer to $h>h_0$ 
that correspond to  $\theta_{\rm s}=0.44$ mrad for each $R_{\rm a}$ indicated.
It is seen that although the channeling peaks are shifted to the right they, simultaneously, 
loose the intensity.
Apart from this, the left peaks become more powerful being centered at $\theta_{\rm s}=0$
due to the increase in the number of particles moving in the forward direction at the expense
of the volume-reflected ones.  
All these modifications can be explained in terms of the qmBC geometry. 

Two panels in Fig. \ref{Figure04.fig} correspond to two sets of $R_{\rm a}$ and $\sigma$.
In each panel, the simulations have been performed for 
different values of the beam center $h$ at the entrance.
Vertical lines in Fig. S1 in SM allow one to compare the $h$ values indicated
with the boundaries $h_0$ and $h_0+\Delta h_{\max}$.

The \textit{left} panel presents a case study in which
$\Delta h_{\max}=160$ $\mu$m is comparable to the beam size
so that for any entrance point within $[h_0, h_0+\Delta h_{\max}]$ a large fraction of the 
particles is not accepted resulting in a noticeable decrease of the right peak.
The curve with $h=400$ $\mu$m corresponds to the case $h=h_0$ when 
half of the beam enters the crystal having $\Delta h < 0$.
In this domain the volume reflection can occur shifting   
the main maximum towards negative angles.
As $h$ increases the numbers of both channeling and volume reflected particles decrease 
leading to the shift of the both maxima to the right as well as to the change in their heights.
At $h=600$ $\mu$m, which corresponds to $\Delta h > \Delta h_{\max}$, most of the beam particles 
do not comply with the channeling condition but experiencing multiple scattering as in amorphous medium.
As a result the main peak becomes more powerful being centered at $\theta_{\rm s}=0$.

To increase the channeling fraction one can rely on a larger value of the anticlastic radius 
and on a narrower beam.
For $R_{\rm a}=300$ cm, Fig. \ref{Figure04.fig} \textit{right},
the quantities $h_0$ and $\Delta h_{\max}$ are 600 and 240 $\mu$m, respectively.
The latter value together with the reduced beam size ($\sigma = 75$ $\mu$m)
suggest that a much bigger fraction of the particles can be accepted 
provided the condition $0 < \Delta h < \Delta h_{\max}-\sigma$ is met.
The best agreement with the experiment has been found for $h=675$  $\mu$m (open circles).
This dependence is shown in Fig. \ref{Figure02.fig} in the form of a histogram.

The quantitative analysis of the angular distribution of ultrarelativistic electrons deflected by oriented qmBCs
presented in our paper demonstrates the good agreement with experimental data reported in \cite{Wienands2015}.
It has been achieved by accounting for (i) the specific geometry of such crystals and their orientation with respect to the
projectile beam and (ii) the realistic beam size and divergence. 
Remaining discrepancies can be attributed to the uncertainty in concrete values of the beam characteristics
and of the entrance coordinate $h$ of the beam center 
as well as to the effects not included into the current simulations 
(e.g., quantum effects in multiple scattering in crystals \cite{TikhomirovPRAB2019}).
It is highly desirable that such information is provided when presenting the experimental data since it allows for
its independent unambiguous theoretical and computational validation. 
Important issue concerns also accurate measurement and computational analysis of the characteristics of 
radiation that accompany passage of ultra-relativistic projectiles through oriented crystals. 
Such knowledge is essential for better planning of accelerator-based experiments and for full 
interpretation of their results. 

The work was supported in part by the DFG Grant (Project No. 413220201) 
and by the H2020 RISE-NLIGHT project (GA 872196).
We acknowledge helpful discussions with Andrea Mazzolari,  Vincenzo Guidi, Hartmut Backe and Werner Lauth.
Frankfurt Center for Scientific Computing (CSC) is acknowledged for providing computer facilities.


\end{document}